\documentclass[journal]{IEEEtran}
\usepackage{cite}
\usepackage{amsmath,amssymb,amsfonts}
\usepackage{algorithmic}
\usepackage{graphicx}
\usepackage{textcomp}
\usepackage{booktabs} 
\usepackage{multirow}
\usepackage{amssymb}

\begin{document}
\title{PCA: Semi-supervised Segmentation with \\Patch Confidence Adversarial Training}
\author{Zihang Xu, Zhenghua Xu, \IEEEmembership{Member IEEE}, Shuo Zhang, Thomas Lukasiewicz  % member or fellow
\thanks{This work was supported by the National Natural Science Foundation of China under the grant 61906063, by the Natural Science Foundation of Hebei Province, China, under the grant F2021202064, by the ``100 Talents Plan'' of Hebei Province, China, under the grant E2019050017, by the Natural Science Foundation of Tianjin City, China, under the grant 19JCQNJC00400, and by the Key Research and Development Project of Hainan Province, China, under the grant ZDYF2022SHFZ015. This work was also partially supported by the AXA Research Fund. (Corresponding author: Zhenghua Xu, e-mail: zhenghua.xu@hebut.edu.cn.)}
\thanks{Zihang Xu, Shuo Zhang and Zhenghua Xu are with State Key Laboratory of Reliability and Intelligence of Electrical Equipment and Tianjin Key Laboratory of Bioelectromagnetic Technology and Intelligent Health, Hebei University of Technology, China.}
\thanks{Thomas Lukasiewicz is with Department of Computer Science, University of Oxford, United Kingdom.}
}

\maketitle

\begin{abstract}
Deep learning based semi-supervised learning (SSL) methods have achieved strong performance in medical image segmentation, which can alleviate doctors’ expensive annotation by utilizing a large amount of unlabeled data. Unlike most existing semi-supervised learning methods, adversarial training based methods distinguish samples from different sources by learning the data distribution of the segmentation map, leading the segmenter to generate more accurate predictions. We argue that the current performance restrictions for such approaches are the problems of feature extraction and learning preference. In this paper, we propose a new semi-supervised adversarial method called Patch Confidence Adversarial Training (PCA) for medical image segmentation. Rather than single scalar classification results or pixel-level confidence maps, our proposed discriminator creates patch confidence maps and classifies them at the scale of the patches. The prediction of unlabeled data learns the pixel structure and context information in each patch to get enough gradient feedback, which aids the discriminator in convergent to an optimal state and improves semi-supervised segmentation performance. Furthermore, at the discriminator's input, we supplement semantic information constraints on images, making it simpler for unlabeled data to fit the expected data distribution. Extensive experiments on the Automated Cardiac Diagnosis Challenge (ACDC) 2017 dataset and the Brain Tumor Segmentation (BraTS) 2019 challenge dataset show that our method outperforms the state-of-the-art semi-supervised methods, which demonstrates its effectiveness for medical image segmentation.
\end{abstract}


\section{Introduction}
\label{sec:introduction}
\IEEEPARstart{S}{egmentation}, identifying interesting regions with anatomical or pathological structures from medical images, is the basic task of medical image analysis, which is of great significance for computer-assisted diagnosis, surgery simulation, and treatment planning. Recently, deep learning methods \cite{1,2} have achieved excellent results in various fields of medical applications, which are trained with various typical segmentation networks in a fully supervised way, (e.g., FCNs \cite{3}, U-Nets \cite{4} and GAN \cite{5}). The success of the deep network model is due to the deep and wide network, which is highly dependent on large-scale and high-quality pixel annotation data. However, the lack of sufficient labeled data has always been a major challenge for medical image segmentation.
\begin{figure}[t]
\includegraphics[width=\columnwidth]{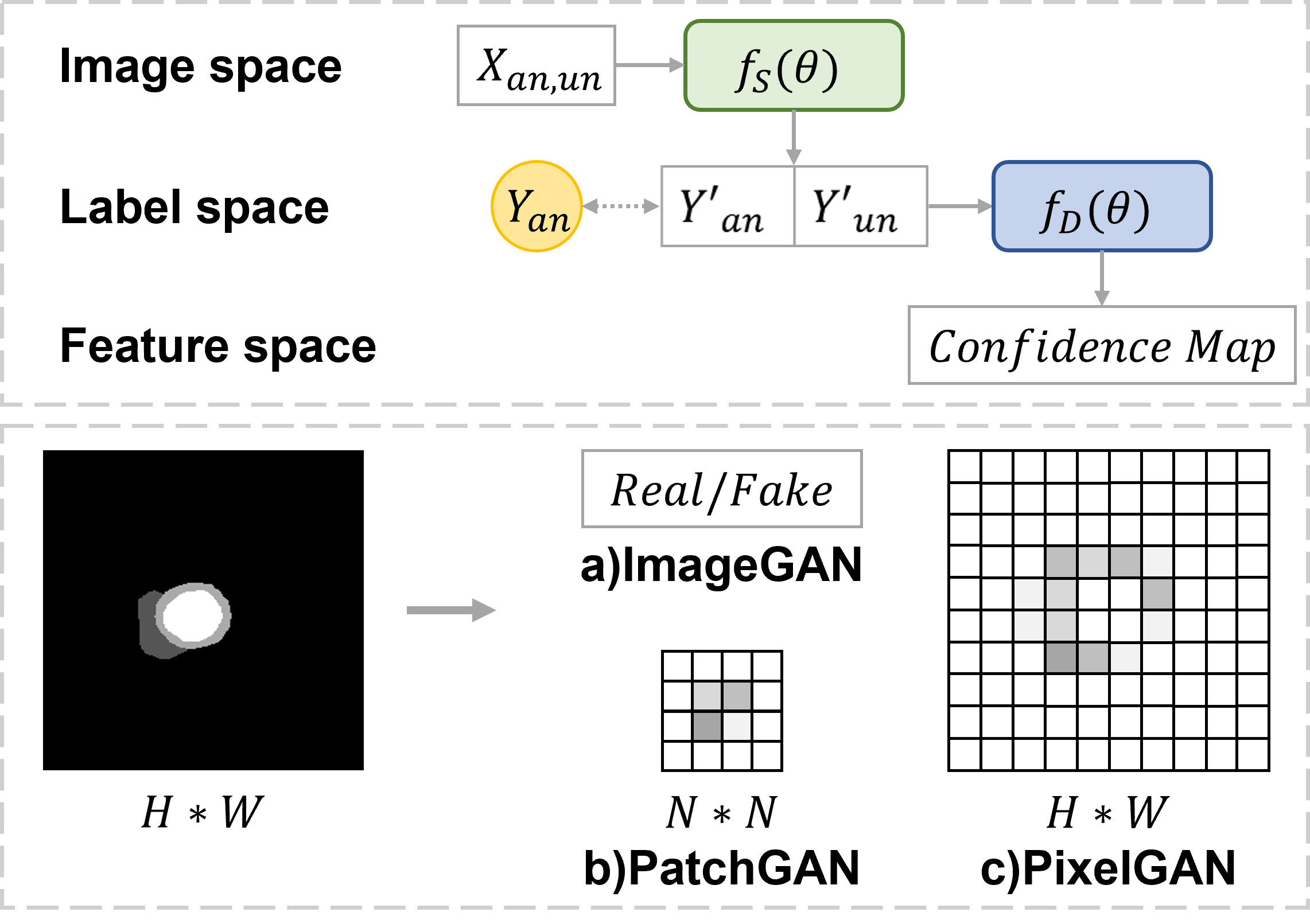}
\caption{Different output styles of general adversarial
training (ImageGAN), our method (PatchGAN), and confidence map based adversarial
training (PixelGAN).
%This is the variety of discriminator from ImageGAN to PatchGAN to PixelGAN, corresponding semi-supervised general adversarial training, our method and semi-supervised confidence map adversarial training.
}\vspace{-1.5em}
\end{figure}

To reduce the annotation burden on doctors, many methods have been proposed for medical image segmentation applications. Considering that unlabeled data is usually abundant and easily available, many researchers focused on implementing segmentation tasks in a semi-supervised learning fashion. The main idea of semi-supervised segmentation is to learn from a limited amount of labeled data and a large amount of unlabeled data to improve the accuracy of segmentation. According to the training manner, common semi-supervised segmentation methods can be divided into self-training \cite{24}, consistency regularization \cite{25}, co-training \cite{26} and adversarial training \cite{27}.

Recently, many researchers have focused on semi-supervised segmentation methods based on adversarial training \cite{28,29,31,34,52,53}, which shows great potential for improving semantic segmentation. Specifically, these methods are inspired by the Generative Adversarial Network (GAN) \cite{5}, which consists of a generator and a discriminator. The generator network uses a semantic segmentation network to generate the probability maps of the semantic labels. The discriminator network generates the probability that the input is real by distinguishing generated samples from target ones. Through the discriminator, the quality of the predicted segmentation map can be effectively assessed. However, the model lacks stability due to the insufficient supervision of the discriminator, which brings challenges to semi-supervised adversarial training. Some recent studies have attempted to improve the performance of the discriminator \cite{28,29,31}. For example, Son \textit{et al.} \cite{28} connected prediction, inverse prediction, and image, forcing the discriminator to learn the mapping relationship between image and prediction, thus alleviating this problem. Zhang \textit{et al.} \cite{29} multiplied the prediction with the image to obtain an image containing only foreground information. Inspired by the attention mechanism \cite{30}, Han \textit{et al.} \cite{31} proposed a dual-attentive-fusion block that has two independent spatial attention paths on the predicted segmentation map and leverages the corresponding original image. Nie \textit{et al.} \cite{52} introduced the concept of a confidence map to supervise the learning of unannotated data.

However, all of the above methods have the following two shortcomings: (i) \textit{Feature mining problem}: For the discriminator, its purpose is to distinguish whether the generated sample is real or not, and its output can represent the probability that the input is real. However, regardless of whether the segmentations generated from the unlabeled data are accurate or not, it is unreasonable to regard all pixels as a negative class, which makes the discriminator difficult to dig out the features that distinguish the distribution of the two data domains. When its output is not a scalar result but a pixel-level confidence map, i.e. classifying each pixel, unlabeled data can use the pseudo label mask generated based on the confidence map to train in a self-learning manner. However, the generator can easily deceive the discriminator by reducing the information entropy of segmentation maps. That means the discriminator doesn't dig out enough features to generate reliable confidence maps to reflect the correct probability of the segmentation result. This motivates us to consider how to generate more reliable evaluation results. (ii) \textit{Learning preference problem}: The original image directly concatenated or multiplied with the segmentation map as the input of the discriminator not only introduces the relevant information of the segmentation map and the input image but also makes the discriminator generate learning preferences. This causes the discriminator to change the optimization goal of training, which affects the accuracy of the segmentation network. From the results, not only the segmentation accuracy has decreased, but also the problem of under-fitting, which is unacceptable in semi-supervised medical image segmentation and may be difficult to apply to clinical practice.

Inspired by these, we propose a patch confidence adversarial training framework for semi-supervised medical image segmentation. We improve the original adversarial training strategy from the input and output of the discriminator, addressing the two problems mentioned above, respectively. First, to better evaluate segmentation, we introduce the idea of PatchGAN \cite{54} . The discriminator classifies each image patch independently, so that it generates a patch confidence map instead of a scalar classification result or pixel-level confidence map. The discriminator penalizes structure at the scale of patches, guiding the generator to optimize for different patch areas. At the same time, with the help of the context information between pixels, the generator is forced to focus on high-frequency features, and it is no longer easy to fool the discriminator, which helps the discriminator converge to an ideal state. In addition, to utilize the image information more effectively, we also add the weighted image and the segmentation map. This operation balances the information contained in the image and segmentation maps. In that case, the discriminator will focus on the relationship between the image and the segmentation, learning how to map unlabeled data to an expected distribution. In adversarial training, improving the performance of the discriminator will help the deep model achieve better segmentation performance.

The major contributions can be summarized as follows:
\begin{itemize}
\item We propose a novel and universal semi-supervised method, namely PCA, for semi-supervised medical image segmentation tasks. The proposed method effectively addresses two important shortfalls, including the feature mining problem and the learning preference problem.
\item Inspired by CGAN \cite{66} and PatchGAN, we propose the patch confidence map and pixel additive blending, making the discriminator easily converge to a desirable status and guiding the prediction of the unlabeled data to fit the expected data distribution, which can significantly improve the performance of semi-supervised segmentation.
\item A large number of experiments on the ACDC 2017 dataset and BraTS 2019 dataset prove that our semi-supervised method is efficient. We found that the designed patch confidence map is more effective than the most advanced adversarial training strategy. In addition, we achieve the best results compared to other state-of-the-art methods in semi-supervised segmentation.
\end{itemize}     
\begin{figure*}[t]
\centering
\includegraphics[width=\textwidth]{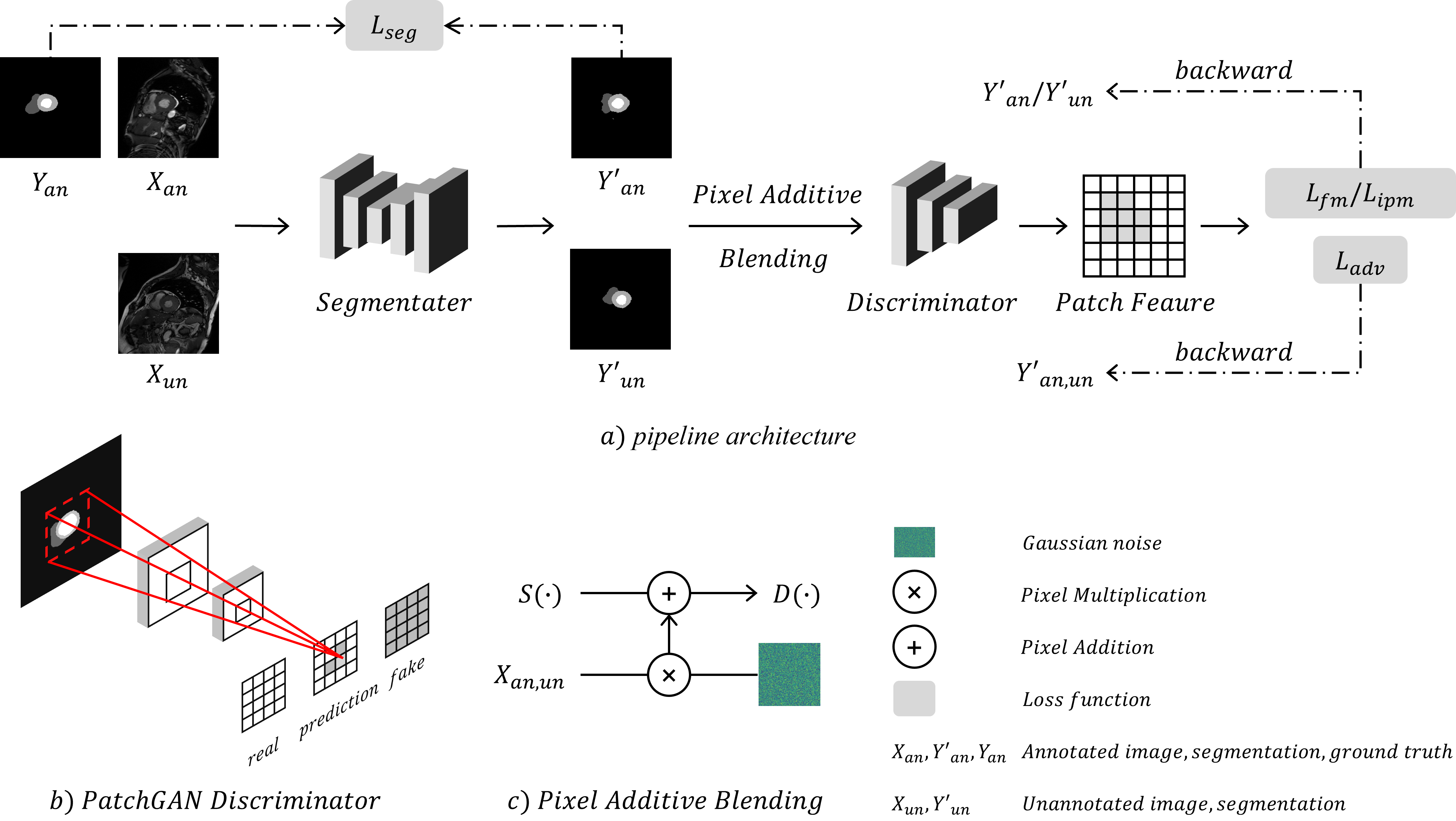}
\caption{a) The schematic view of our proposed patch confidence adversarial training model (take cardiac image as an example). b) Each value of the output matrix from PatchGAN discriminator represents the probability of whether the corresponding segmentation patch is real or fake. c) The process of pixel additive blending calculation.}\vspace{-1.5em}
\label{fig20}
\end{figure*}

\section{RELATED WORKS}
As a basic task, medical image segmentation is of great significance in many biomedical applications. At present, almost all segmentation frameworks are based on deep learning, which has achieved impressive performance improvements on various medical image segmentation tasks and set the new state-of-the-art. However, there is a lack of a large amount of data annotation in the medical field, which limits the performance of the model. In this section, we will focus on reviewing related methods and the latest developments in semi-supervised learning, and then discuss semi-supervised methods based on adversarial training, which are most relevant to our work.

\subsection{Semi-supervised segmentation for medical images}
In semi-supervised medical image segmentation tasks, it is usually assumed that only a small part of the training images have complete pixel-level annotations, but there are also a large number of unlabeled images that can be used to improve the performance of the model. Since unlabeled data does not need to be manually labeled by the doctor, unlabeled data can be used at a low cost to improve performance. The main challenge in this scenario is how to effectively use a large amount of unlabeled data, which is also the main difference between different methods.

Recently, almost all semi-supervised medical image segmentation frameworks are based on deep learning. For example, Bai et al. \cite{11} proposed an iterative method for heart segmentation of MR images, using pseudo-labels generated by network prediction to update network parameters. Feng \textit{et al.} \cite{12} improved the training strategy based on the work of Bai \textit{et al.} \cite{11}. Only part of the reasonable segmentation maps predicted from the unannotated samples was progressively combined with the annotated samples to improve the training procedure. Li \textit{et al.} \cite{13} proposed a transformation consistent self-integration model for semi-supervised skin lesion segmentation based on the $\pi$ model. Inspired by the Mean Teacher model, Yu \textit{et al.} \cite{14} proposed a semi-supervised framework for uncertainty perception to segment the left atrium from 3D MR images. Cao \textit{et al.} \cite{15} extended the uncertain time model for semi-supervised ABUS quality segmentation. However, such methods do not consider the quality of the predicted segmentation map, which may introduce misinformation into the segmentation network. Adversarial training introduces a discrimination network to evaluate the predicted segmentation map, which is simple and effective for semi-supervised medical image segmentation.

\subsection{Adversarial training for Semi-Supervised Segmentation}
The Generative Adversarial Network (GAN) is proposed by Goodfellow et al. \cite{5}. It generates samples by optimizing the adversarial game between the discriminator and the generator. When applying adversarial learning to semi-supervised segmentation tasks, the model usually uses two networks: a segmentation network (generator) for image segmentation and a discriminant network (discriminator) to identify whether the sample is extracted from real data or generated by the generator. Generally, there are three popular strategies in the medical image analysis community based on adversarial training and semi-supervised methods, including generative models, confidence maps, and segmentation evaluation. For instance, Sedai et al. \cite{32} introduced a generative model (VAE \cite{33}) for optic cup (OC) segmentation from retinal fundus images. They used VAE to learn feature embedding from unlabeled images and then combined the feature embedding with a segmentation autoencoder trained on labeled images to perform pixel segmentation on the cup region. Lahiri \textit{et al.} \cite{53} deployed unannotated images to generate the fake samples to increase the amount of the training dataset. However, the model lacks stability due to the insufficient supervision of the discriminator.

To improve the performance of the discriminator, a simple strategy is inspired by conditional GAN, where the discriminator is conditioned on the input image to classify whether the segmentation map is real or fake. Son \textit{et al.} \cite{28} connected prediction, inverse prediction, and image, and distinguished good or poor segmentation results by finding the mapping relationship between image and prediction. Zhang \textit{et al.} \cite{29} multiplied the prediction with the image to judge directly whether the target is accurately segmented. Han \textit{et al.} \cite{31} introduced a dual attention fusion block based on connection, extracting geometric level and intensity level information, thus digging for more relevant features. Such works explore the correlations between the segmentation and input image for evaluating the segmentation quality. However, regardless of whether the segmentation and the corresponding image are related or not, it is unreasonable to regard all pixels in the generated samples as negative samples, which is too abstract for the discriminator. As a consequence, the generated evaluation score does not reflect the correct probability of the segmentation result, and the discriminator cannot contribute to instructing the segmentation task.

Nie \textit{et al.} \cite{34} adopted another strategy, using a discriminator to output the confidence map and choose high-confidence regions to obtain the ground truth, then updating the segmentation network in a self-learning manner. The main limitation of this method is that a confidence threshold must be provided, the value of which will affect performance. What's more, if the discriminator cannot distinguish between good and bad segmentation, poor confidence maps may reduce the performance of the entire adversarial training. In particular, for medical images, the presence of speckle noise may affect the confidence map. The unstable confidence map may also affect the entire learning process and lead to unsatisfactory segmentation results for unlabeled data.

At the input of the discriminator, we design a new combination of conditional GAN, called pixel additive blending. At the output of the discriminator, different from the confidence map and single scalar output, we use PatchGAN to output an $N\times N$ array to evaluate the good or bad segmentation results in a region, rather than the entire image or individual pixels. Such improvements effectively solve the feature mining and learning preference problems that exist in other methods.

\section{Method}
Fig. 2 shows a schematic illustration of our patch confidence adversarial training model (PCA) for semi-supervised segmentation. The PCA framework consists of a segmentation network and a discrimination network. First, the segmentation network takes input data and produces the corresponding segmentation probability maps. Then, the segmentation probability map and the image weighted combination are used as the inputs of the conditional discriminator. Subsequently, the discriminator distinguishes the data distribution of new images generated based on different maps, including the ground truth from the annotation data and the segmentation probability maps from all the data. The discriminator tries to classify if each $N\times N$ patch in a segmentation map is real or fake.

\subsection{General training strategy}
To define the loss function, the symbols utilized is first enlisted. The training set consists of $M+N$ inputs, including $M$ labeled inputs and $N$ unlabeled inputs. Let $A=\{(x_i^a,y_i^a)\}_{i=1}^M$ be a labeled set with $M$ samples and $U=\{x_i^u\}_{i=1}^N$ be a set with $N$ samples. The segmentation network and the discrimination network are represented by $S(\cdot)$ and $D(\cdot)$. 

In our proposed PCA, the segmentation network is trained by minimizing the following loss function $L_S$:
\begin{equation}
L_S(A,U;\theta_S)=L_{seg}+\lambda_{adv} L_{adv},
\end{equation}
where $\theta_S$ represents the parameters for the segmentation network, $L_{seg}$ and $L_{adv}$ represent the supervised segmentation and adversarial loss, respectively. $\lambda_{adv}$ refers to the weight of adversarial learning.

The loss function $L_{seg}$ is used to determine whether the segmentation probability map generated by the input annotation data is close to ground truth, which is expressed as:
\begin{equation}
L_{seg}(S(x_i^a),y_i^a)=
0.5\ast L_{bce}+0.5\ast L_{dice},
\end{equation}
\begin{equation}
\begin{split}
L_{bce}(S(x_i^a),y_i^a)=&-y_i^a\cdot{\rm log}(S(x_i^a))\\
&-(1-y_i^a)\cdot{\rm log}(1-S(x_i^a))),
\end{split}
\end{equation}
\begin{equation}
L_{dice}(S(x_i^a),y_i^a)=1-\frac{2\ast \sum S(x_i^a)\ast y_i^a}{\sum S(x_i^a)+\sum y_i^a},
\end{equation}
where $L_{bce}$ is constrained by the standard Binary Cross Entropy (BCE) loss function, $L_{dice}$ is constrained by the standard DICE loss function.

The general semi-supervised adversarial training objective loss function $L$ is defined as:
\begin{equation}
\begin{split}
\min \limits_{\theta_S}\max \limits_{\theta_D} L(\theta_S,\theta_D)=
{\rm E}_{x_i\sim P_{A}}[{-\rm log}D(S(x_i))]&\\
+{\rm E}_{x_i\sim P_{U}}[{-\rm log}(1-D(S(x_i)))]&,
\end{split}
\end{equation}
where $x_i$ is the input image from data distribution $P_{A+U}$, and $\theta_D$ represents the parameters for the discrimination network. As in GAN \cite{5}, when updating the segmentation network, we replace the ${\rm log}(1-D(x_i))$ by ${-\rm log}D(x_i)$.

\subsection{Patch confidence map}
The choice of the discriminator differs due to the different output sizes at which the decision is made. In this work, we explored several models for the discriminators with various output sizes, as done in the previous work, corresponding to several adversarial training strategies. Notably, our $D(\cdot)$ produces a $N\times N$ matrix as a result.

According to the general adversarial training strategy, the image-level discriminator’s purpose is to determine whether the input sample is real. The last layer of the general discriminator outputs a scalar value, which is a weighted value of the whole image but cannot reflect the local features of the image. It is hard to train for tasks requiring high precision, such as image segmentation. Especially in semi-supervised segmentation tasks, if the discriminator treats predictions from unlabeled data as negative samples and only learns global features, it won’t be easy to mine enough features to guide the segmenter to generate more accurate predictions.

The pixel-level discriminator generates the confidence map to determine whether each pixel in the input sample is real or not. Afterwards, unlabeled data is trained by generating reliable pseudo-label masks from the confidence map. This requires the confidence map to infer sufficiently close regions from the ground truth distribution to find more confident pixels. However, the confidence map generated by the current method cannot reflect the correct probability of the segmentation result, which will affect the performance of the segmentation network and even cause a negative transfer.

Between the extremes, it is also possible to set the receptive field to a $K\times K$ patch where the decision can be given at the patch-level (PatchGAN). The final output of the patch-based discriminator is a matrix with a size of $N\times N$. Each of the items in the matrix represents a local region in the image. The patch-based discriminator tries to classify whether each $N\times N$ patch in the image is real or fake, which achieves the extraction and characterization of local image features and is conducive to the generation of high-precision images. Compared with image-level (ImageGAN) or pixel-level (PixelGAN) adversarial learning, PatchGAN has the ability to capture the local statistics of the output space and guide the segmentation network to focus on the local structure similarity in the image patches. We achieve this variation in patch size by adjusting the depth of the GAN discriminator.

In the early stages of the training process, the positive and negative sample distributions may not overlap each other, where the discriminator network can easily distinguish them. To stabilize the training of the semi-supervised framework, we directly align the features learned by the discriminator network to the segmentation network and introduce the feature matching \cite{67} loss $L_{fm}$. It aims to minimize the discrepancy in feature statistics between the generating samples and the ground truth. It is calculated by a mean squared error (MSE) loss:
\begin{equation}
L_{fm}(x_i,y_i)={\rm E}_{x_i,y_i\sim P_{A}}||D(S(x_i))-D(y_i)||^2,
\end{equation}
where $x_i$ and $y_i$ are sampled from the labeled set $A$.

For unlabeled data, it is impossible to align ground truth features directly. Inspired by the integrated probability metric \cite{68}, we choose the intermediate layer of the discriminative network as the mapping function and calculate the mean value to measure the distance between the distributions of positive and negative samples. We believe the two distributions are the same when the feature centres match (i.e. the difference in means is minimal). $L_{ipm}$ is used for unlabeled prediction, forcing the segmentation network to use a reasonable solution without label information. $L_{ipm}$ displays in Eq. (7):
\begin{equation}
L_{ipm}(x_i,y_i)=||{\rm E}_{x_i\sim P_{U}}D(S(x_i))-{\rm E}_{y_i\sim P_{A}}D(y_i)||^2,
\end{equation}
where $x_i$ represents unlabeled images sampled from unlabeled set $U$, and $y_i$ represents the ground truth sampled from labeled set $A$.
The final training objective $L_S$ is as follows:
\begin{equation}
L_S=L_{seg}+\lambda_{adv}L_{adv}+\lambda_{fea}L_{fea}+\lambda_{ipm}L_{ipm},
\end{equation}
where $\lambda_{fea}$ and $\lambda_{ipm}$ are the corresponding weights.

\subsection{Pixel additive blending}
The simplest input form can be just the segmentation probability maps \cite{46,47,48}, which allow the discriminator to learn useful shape properties of the objects, thereby evaluating the segmentation result quality. However, in this form, the discriminator has weaker discriminability, which can easily cause an over-fitting problem.

In conditional GAN (CGAN), the discrimination model explores the relationship between the segmentation probability map and the image, to enhance the discriminative ability. In that case, the discriminator might generate the probability values by distinguishing between labeled and unlabeled images without learning segmentation information. So the key is how to encode the correlation information to construct the effective input of the discriminator. Below, we will discuss several methods for combining the original image enhancement segmentation results.

The segmentation result can be directly concatenated to the original image \cite{28}. Since the discrimination network has separate model parameters for handling information from the segmentation maps and from the original image, the discriminator may focus on learning the features of the original image to give a judgment score, which leads to the learning preference problem.

Element-wise multiplication \cite{29} forces the discriminator to learn about the association information between the segmentation maps and the original images, which ensures that the segmentation maps are meaningful in adversarial training. However, element-wise multiplication ignores the correlation between the background and the target. The discriminator will focus on the distinction of some uncomplicated categories, which is not conducive to improving its ability.

The attention block \cite{31} is utilized to generate the attentive rated maps from the segmentation maps and the corresponding input images. The image's features are multiplied by the attentive rating map in an element-by-element manner to generate the attentive feature map. However, the attention method does not deal with all the details of the image accurately, introducing a lot of noisy information, which is usually fatal for medical image segmentation tasks.

\begin{figure}[t]
\includegraphics[width=\columnwidth]{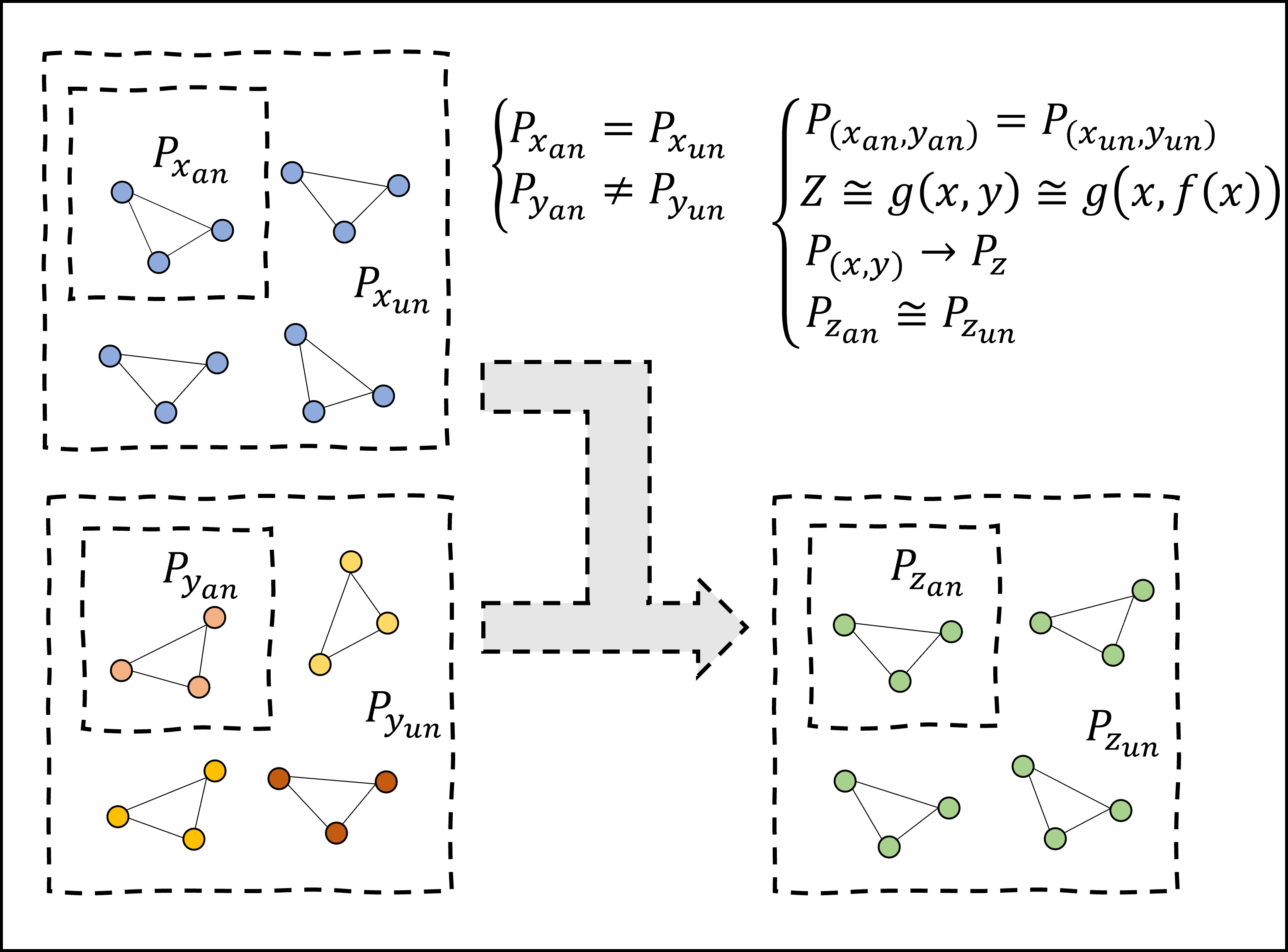}
\caption{Using the condition that the labeled data and unlabeled data are identically distributed, the input and output are coupled, and the obtained joint probability is approximately identically distributed.}\vspace{-1.5em}
\end{figure}

To alleviate the influence of learning preference on network optimization and improve feature mining ability, we introduce a novel conditional GAN, called pixel additive blending, where the discriminator is conditioned on the input image $x$. The main idea is to fit the joint probability of the image and the segmentation to narrow the gap between the labeled data pair and the unlabeled data pair, thereby helping the discriminator to better learn the correlation between the image and the segmentation. This is because semi-supervised learning assumes that the labeled data and unlabeled data samples belong to the same distribution, but there is a gap between their corresponding segmentations. If the joint probability distribution of the image and the segmentation can be fitted, then the joint probability distribution of the unlabeled data will be the same as that of the labeled data. close. Figure 3 shows the process information.

Our solution is to assign a Gaussian noise with a small value to the image and add it to the segmentation map pixel by pixel. It is worth mentioning that here we use the segmenter's predictions as negative samples and the ground truth of the labeled data as positive samples. It is to avoid the discriminator to complete the classification task by distinguishing between labeled and unlabeled images, which will deviate from the original purpose. Thus, the objective function becomes:
\begin{equation}
\begin{split}
\min \limits_{\theta_S}\max \limits_{\theta_D} L(\theta_S,\theta_D)=
{\rm E}_{x_i,y_i\sim P_{A}}[{-\rm log}D(Z(x_i,y_i))]&\\
+{\rm E}_{x_i\sim P_{A+U}}[{-\rm log}(1-D(Z(x_i,S(x_i))))]&.
\end{split}
\end{equation}
Particularly, the pixel additive blending $Z(\cdot)$ is expressed as follows:
\begin{equation}
Z(x_i,y_i)=y_i+\lambda_{noise}x_i\cdot noise,
\end{equation}
where $\lambda_{noise}$ and $noise$ represent the weight coefficient of the image and Gaussian noise, respectively. All $D(x)$ are replaced by $D(Z(x_i,y_i))$, when introducing the pixel additive blending.

Here are three reasons: Firstly, pixel-addition ensures both the segmentation map and the image are used in the discriminator's decision-making and in the adversarial training process. Also, the randomness of Gaussian noise reduces the correlation between image pixels, forcing the discriminator to learn the correlation between segmentation and image. In addition, we believe that the output is less complex than the input in the semi-supervised segmentation task. That means the segmentation map may have very little influence on both the decision-making process of the discriminator and the parameter updates of the segmenter. In order to reduce this bias, we must limit the influence of image information. Here we assign a small weight to the image.

\subsection{Network architecture}
For the segmentation network, we follow the spirit of U-Net where skip connections are added between the down-sampling path and the up-sampling path to save low-level information. This skip-connection is crucial to segmentation tasks as the initial feature maps maintain low-level features such as edges and blobs that can be properly exploited for accurate segmentation. In addition, the U-Net model utilizes bi-linear interpolation to expand the feature maps.

The proposed discrimination network generally follows the architecture of PatchGAN \cite{54}. The network contains three convolutional layers with a kernel size of $4\times 4$ and a stride of $2\times2$. The channel numbers of the three convolutional layers are 32, 64, and 1, respectively. The activation function following each convolutional layer is LeakyReLU with an alpha value of 0.2, except for the last one using the Sigmoid function. The output size ($m\times n$) the patch-based discriminator is $32\times32$, in which one pixel corresponds to a patch of size $22\times22$ in the input probability maps. Each patch is classified into real (1) or fake (0) through the discriminator. We employ this adversarial learning strategy to force each generated patch in the prediction of unlabeled data to be similar to the patch of labeled data.

\section{Experiments}

\subsection{Dataset}
We evaluated our proposed semi-supervised segmentation method on two typical medical images, including cardiac images and brain tumor images.

\textbf{Cardiac image dataset.}
This dataset consists of 200 MRI scans from 100 patients \cite{62} for training and 50 patients for testing. Three cardiac regions are labeled in the ground truth: left ventricle (LV), right ventricle (RV), and myocardium (Myo). For a fair comparison, we only selected the training set in our experiments and divided the data set into the training, validation, and testing sets, respectively, containing 70, 10, and 20 patients’ data. Slices within 3D\-MRI scans were considered 2D images, which were fed as input to the network.

\textbf{Brain tumor image dataset.}
The brain tumor image dataset comes from the Brain Tumor Segmentation (BraTS) 2019 challenge. The released BraTs dataset contains 335 3D cases. The experiment of whole brain tumor segmentation was performed using the multi-modal MRI data from the BraTS 2019 challenge \cite{63,64,65}. The entire dataset contains multi-institutional preoperative MRI of 335 glioma patients (259 HGG and 76 LGG), where each patient has four modalities of MRI scans with neuroradiologist-examined pixel-wise labels. Here, we use the T1-CE modal of HGG for whole tumor segmentation since this modality can better manifest malignant tumors. In our experiments, the MRI scans are normalized to zero mean and unit variance. We randomly selected 207 samples for training and 52 for validation. We slice 3D imags into 2D images and crop them to $160\times 160$.
\begin{table*}[!ht]
\caption{Comparisons with state-of-the-art methods on the ACDC2017 dataset}
\centering
\scalebox{1}{
\begin{tabular}{c|cc|cccc|cccc}
\toprule
\multirow{2}*{Method}&\multicolumn{2}{c}{Scans used}&\multicolumn{4}{|c}{Area Overlap}&\multicolumn{4}{|c}{Boundary Error}\\ \cline{2-11}
&An.&Un.&DSC(\%)$\uparrow$&Imp.&JA(\%)$\uparrow$&Imp.&95HD(voxel)$\downarrow$&Imp.&ASD(voxel)$\downarrow$&Imp.\\
\hline
Baseline&7(10\%)&0&77.34&-&66.20&-&9.18&-&2.45&-\\
\hline
GAN \cite{28}&\multirow{14}*{7(10\%)}&\multirow{14}*{63(90\%)}&83.28&5.94\%&72.84&6.64\%&9.84&-0.66&2.78&-0.33\\
DAN \cite{29}&&&79.33&1.99\%&67.93&1.73\%&11.85&-2.67&3.08&-0.63\\
BUS-GAN \cite{58}&&&81.80&4.46\%&71.20&5.00\%&6.11&3.07&2.21&0.24\\
ASDNet \cite{52}&&&82.89&5.55\%&71.98&5.78\%&16.50&-7.32&4.13&-1.68\\
SASSNet \cite{50}&&&84.14&6.80\%&74.09&7.89\%&5.03&4.15&1.40&1.05\\
MT \cite{39}&&&80.40&3.06\%&69.28&3.08\%&10.05&-0.87&2.65&-0.20\\
ICT \cite{38}&&&83.54&6.20\%&72.84&6.64\%&7.58&1.60&2.27&0.18\\
UAMT \cite{14}&&&81.58&4.24\%&70.48&4.28\%&12.35&-3.17&3.62&-1.17\\
CCT \cite{59}&&&83.34&6.00\%&72.84&6.64\%&7.07&2.11&2.18&0.27\\
DTC \cite{57}&&&82.71&5.37\%&72.14&5.94\%&11.31&-2.13&2.99&-0.54\\
CPS \cite{60}&&&85.32&7.98\%&75.42&9.22\%&6.64&2.54&1.98&0.47\\
UPRC \cite{56}&&&81.77&4.43\%&70.85&4.65\%&5.04&4.14&1.41&1.04\\
PCA(w/o CGAN)&&&86.39&9.05\%&76.91&10.71\%&4.63&4.55&\textbf{1.20}&\textbf{1.25}\\
PCA(w/ CGAN)&&&\textbf{87.33}&\textbf{9.99\%}&\textbf{78.31}&\textbf{12.11\%}&\textbf{3.10}&\textbf{6.08}&1.35&1.10\\
\hline
Baseline&14(20\%)&0&83.69&-&74.00&-&6.63&-&1.74&-\\
\hline
GAN \cite{28}&\multirow{14}*{14(20\%)}&\multirow{14}*{56(80\%)}&84.52&0.83\%&74.92&0.92\%&10.51&-3.20&2.64&-0.29\\
DAN \cite{29}&&&86.18&2.49\%&76.71&2.71\%&9.23&-1.92&2.38&-0.03\\
BUS-GAN \cite{58}&&&85.01&1.32\%&75.83&1.83\%&6.64&0.67&1.76&0.59\\
ASDNet \cite{52}&&&84.18&0.49\%&74.01&0.01\%&6.68&0.63&2.10&0.25\\
SASSNet \cite{50}&&&85.99&2.30\%&76.63&2.63\%&5.32&1.99&1.47&0.88\\
MT\cite{39}&&&85.58&1.89\%&76.38&2.38\%&4.89&2.42&1.60&0.75\\
ICT \cite{38}&&&85.25&1.56\%&75.71&1.71\%&7.66&-0.35&2.32&0.03\\
UAMT \cite{14}&&&85.87&0.72\%&76.78&1.30\%&5.06&1.14&1.54&0.58\\
CCT \cite{59}&&&86.09&2.40\%&77.05&3.05\%&7.01&0.30&1.98&0.37\\
DTC \cite{57}&&&86.28&1.13\%&77.03&1.55\%&6.14&0.06&2.11&0.01\\
CPS \cite{60}&&&87.38&3.69\%&78.61&4.61\%&6.06&1.25&1.69&0.66\\
UPRC \cite{56}&&&85.07&-0.08\%&75.61&0.13\%&6.26&-0.06&1.77&0.35\\
PCA(w/o CGAN)&&&87.86&4.17\%&79.13&5.13\%&5.10&2.21&1.51&0.84\\
PCA(w/ CGAN)&&&\textbf{88.09}&\textbf{4.40\%}&\textbf{79.44}&\textbf{5.44\%}&\textbf{2.91}&\textbf{4.40}&\textbf{0.98}&\textbf{1.37}\\
\hline
Upper bound&70(100\%)&0&91.65&-&84.93&-&1.89&-&0.56&-\\
\bottomrule 
\end{tabular}}
\end{table*}

\subsection{Implementation Details}
We implement our framework in PyTorch with 2 Nvidia 2080Ti GPUs. On the ACDC dataset, all settings followed the public benchmark Luo \textit{et al.} \cite{55} for fair comparisons. We normalized the samples as a 0–1 range and used data augmentaition for the random rotation and flip operations. All the 2D patches were interpolated to a size of $256\times 256$ and randomly extracted. The batch size was set at 24, including 12 labeled samples and 12 unlabeled samples. The model was trained via 30K iterations. To train the segmentation network and discrimination network, an SGD and an Adam optimizer were employed to minimize $l_s$ and $l_d$. The initial learning rates were set to 1e-2 and 1e-4, respectively, and decayed according to the equation $lr=(1-\frac{iterations}{iterartions_{total}})^{0.9}$. During the testing time, we also interpolated the output results to $256\times 256$ as input and then restored them to their original size. For the brain tumor image dataset, we normalized the samples as zero mean and unit variance. The batch size was set at 30. The initial learning rates were set at 2e-2 and 1e-4. The model was trained over 100 epochs. After obtaining the segmentation probability map from the segmentation network, we apply thresholding with 0.5 to generate a binary segmentation result. $\lambda_{adv}$,$\lambda_{fea}$,$\lambda_{ipm}$ and $\lambda_{noise}$ were set to 0.1, 1, 0.1 and 0.001.

\subsection{Evaluation metrics}
Following Luo \textit{et al.} \cite{55}, we adopt four metrics, including dice similarity coefficient (DSC), Jaccard (JA), the 95\% Hausdorff Distance (95HD), and the average surface distance (ASD). DSC and JA were employed to examine the overlap areas between the two comparisons. 95HD and ASD were exploited to measure the euclidean distance between a computer-identified lesion boundary and the boundary determined by physicians. Higher DSC and JA, along with lower 95HD and ASD, correspond to the higher similarity between the two compared regions.

\subsection{Experiments on Cardiac Image Data Set}

\begin{table*}[ht]
\caption{Discussion of different proporation of data to train our model on the ACDC2017 dataset}
\centering
\begin{tabular}{c|c|cccc|cccc}
\toprule
\multirow{2}*{Label/Unlabel}&\multirow{2}*{Model}&\multicolumn{4}{|c}{Area Overlap}&\multicolumn{4}{|c}{Boundary Error}\\\cline{3-10}
&&DSC(\%)$\uparrow$&Imp.&JA(\%)$\uparrow$&Imp.&95HD(voxel)$\downarrow$&Imp.&ASD(voxel)$\downarrow$&Imp.\\
\hline
\multirow{3}*{3/67}&Baseline&48.09&-&36.54&-&50.04&-&21.03&-\\
&PCA(w/o CGAN)&56.98&8.89\%&46.43&9.89\%&\textbf{14.15}&\textbf{35.89}&7.51&13.52\\
&PCA(w/ CGAN)&\textbf{66.15}&\textbf{18.06\%}&\textbf{54.70}&\textbf{18.16\%}&15.22&22.74&\textbf{5.42}&\textbf{15.61}\\
% \hline
\multirow{3}*{7/63}&Baseline&77.34&-&66.20&-&9.18&-&2.45&-\\
&PCA(w/o CGAN)&86.39&9.05\%&76.91&10.71\%&4.63&4.55&\textbf{1.20}&\textbf{1.25}\\
&PCA(w/ CGAN)&\textbf{87.33}&\textbf{9.99\%}&\textbf{78.31}&\textbf{12.11\%}&\textbf{3.10}&\textbf{6.08}&1.35&1.10\\
% \hline
\multirow{3}*{14/56}&Baseline&83.69&-&74.00&-&7.31&-&2.35&-\\
&PCA(w/o CGAN)&87.86&4.17\%&79.13&5.13\%&5.10&2.21&1.51&0.84\\
&PCA(w/ CGAN)&\textbf{88.09}&\textbf{4.40\%}&\textbf{79.44}&\textbf{5.44\%}&\textbf{2.91}&\textbf{4.40}&\textbf{0.98}&\textbf{1.37}\\
\hline
70/0&Baseline&91.65&-&84.93&-&1.89&-&0.56&-\\
\bottomrule 
\end{tabular}
\end{table*}

\subsubsection{Comparison with other semi-supervised methods}
We compare our method with the latest semi-supervised learning methods, including GAN-based methods \cite{28,29,58,52,50} and current state-of-the-art semi-supervised methods \cite{14,59,57,60,56,38,39}. We implemented these methods and conducted comparative experiments on the public dataset ACDC 2017. The reported results in Table I are the average performance of four classes on the test set. In each semi-supervised setting, we list the performance of the fully-supervised baseline, adversarial training methods, the latest semi-supervised methods, and our methods in turn. We train U-Net with 100\%, 20\% and 10\% of training data as upper bounds and the two baselines. The GAN-based methods perform better compared to the baseline overall, showing the effectiveness of GAN-based methods for semi-supervised segmentation. However, we can notice that the performance of ASDNet and BUS-GAN is similar to that of GAN, indicating that the application of the confidence map and segmentation evaluation are still challenging for semi-supervised segmentation. Compared to GAN-based methods, consistency-based methods achieve comparable performance, demonstrating them effectively utilizing unlabeled data. For example, under the 10\% setting, UAMT achieved a 4.24\% improvement in the DSC indicator. Obviously, compared with other methods, the model we proposed obtained the best quantitative results for DSC and JA performance in each semi-supervised setting. Specifically, we achieves an average DSC and JA improvement of 9.99\% and 12.11\% or 4.40\% and 5.44\% than the fully-supervised baseline trained with 10\% or 20\% labeled data. In addition, by exploiting the unlabeled data effectively, our model almost always obtains the lowest standard values of boundary error (i.e., 95HD and ASD). Fig. 4 shows the visualization of all methods. We can see that these methods often perform well for lesions. However, for the blurry lesions, our method can predict them better than the other methods. Compared with the most advanced semi-supervised methods, our results can describe the target boundary more accurately and retain more details, which also proves the effectiveness of our method.

\subsubsection{Different ratio of labeled data}
To further verify the influence of different percentages of labeled data on performance for our method, a different number of labeled images were selected from the training set. We chose three settings of 3, 7 and 14 cases, which are 4\%, 10\% and 20\% of labeled training data. Table II shows the segmentation results using different numbers of labeled and unlabeled images. All PCA models perform better than the corresponding supervision methods, which shows that our method effectively utilizes unlabeled data and promotes performance. Our method can achieve a substantial surpass of 2.4\% DSC with 10\% labeled scans compared with the baseline with 20\% labeled scans, demonstrating the significant advantage of our method under a small-scale labeled dataset. When the number of labeled data is small (i.e. labeled data = 3), our method obtains a substantial increase (18.06\% DSC and 18.16\% JA) in accuracy over the fully supervised method (48.09\% DSC and 36.54\% JA). What‘s more, our method achieves a large reduction (about 70\% 95HD and 75\% ASD) in boundary error. Both improvements indicate that our proposed method has broad potential for further clinical applications. It can also be noticed that the performance of all methods increases slowly with the increase of labeled data, which illustrates that the performance of models tends to converge as labeled data increases.

\subsubsection{Ablation study}
We propose an ablation study to measure the contribution of different components of the method, and the results are shown in Table III. The abbreviations “Im.”, “Pa.” and “Pi.” stand for ImageGAN, PatchGAN and PixelGAN. Our framework contains two main components: CGAN and PatchGAN. To investigate the effectiveness of each component, we performed an ablation study by adding the two components to the baseline one by one. In addition, we also explore the impact of a variety of discriminators, including PixelGAN and ImageGAN. The experiments are conducted in the setting of three labeled datasets. The results of different settings are presented in Table III. The first line is the supervised baseline model, which was trained with only 3 labels. First, we add the unlabeled data and the adversarial loss. Our method achieves the best results (56.98\% DSC) among the three, surpassing the baseline by 8.89\%. Other adversarial methods are also better than the baseline, showing the effectiveness of the GAN framework. Then, we explore the infulence of the CGAN strategy. From the results, we can observe that conditonal image constraint significantly improves segmentation performance by 63.81\%. We believe it is because the image-segmentation association plays a key role in the few annotations. Finally, with the joint learning of CGAN and PatchGAN, the performance of our framework is further promoted to the state-of-the-art, surpassing the supervised baseline by 18.06\% DSC. It is observed that applying conditonal image constraint alone contributes more to the model's performance.

\begin{table}[!t]
\caption{Ablation studies of our proposed methods on the ACDC2017 dataset}
\setlength{\tabcolsep}{1pt}
\centering
\begin{tabular}{ccccc|cc|cc}
\toprule
\multicolumn{5}{c|}{Setting}&\multicolumn{2}{c|}{Scans used}&\multicolumn{2}{c}{Metrics}\\
\hline
UNet&Im.&Pa.&Pi.&CGAN&An.&Un.&DSC\_Mean&Imp.\\
\hline
\checkmark&&&&&\multirow{6}*{3(4\%)}&\multirow{6}*{67(96\%)}&48.09&-\\
\checkmark&\checkmark&&&&&&54.43&6.34\%\\
\checkmark&&\checkmark&&&&&56.98&8.89\%\\
\checkmark&&&\checkmark&&&&54.49&6.40\%\\
\checkmark&\checkmark&&&\checkmark&&&63.81&15.72\%\\
\checkmark&&\checkmark&&\checkmark&&&\textbf{66.15}&\textbf{18.06\%}\\
\bottomrule 
\end{tabular}
\end{table}

\begin{table*}[!t]
\caption{Comparisons with state-of-the-art methods on the BraTs2019 dataset}
\centering
\begin{tabular}{c|cc|c|c|c|c|c|c|c|c}
\toprule
\multirow{2}*{Method}&\multicolumn{2}{c}{Scans used}&\multicolumn{8}{|c}{Metrics}\\\cline{2-11}
&An.&Un.&DSC\_WT(\%)&Imp.&DSC\_TC(\%)&Imp.&DSC\_ET(\%)&Imp.&DSC\_Mean(\%)&Imp.\\
\hline
Baseline&11(5\%)&0&38.84&-&35.69&-&37.47&-&37.33&-\\
\hline
GAN\cite{28}&\multirow{14}*{11(5\%)}&\multirow{14}*{196(95\%)}&46.50&7.66\%&52.85&17.16\%&50.77&13.30\%&50.04&12.71\%\\
DAN \cite{29}&&&42.50&3.66\%&47.92&12.23\%&47.07&9.60\%&45.83&8.50\%\\
BUS-GAN \cite{58}&&&40.25&1.41\%&44.64&8.95\%&44.95&7.48\%&43.28&5.95\%\\
ASDNet \cite{52}&&&39.22&0.38\%&42.61&6.92\%&41.98&4.51\%&41.27&3.94\%\\
SSASNet \cite{50}&&&39.45&0.61\%&49.12&4.49\%&42.37&4.90\%&40.67&3.33\%\\
MT\cite{39}&&&40.27&1.43\%&43.43&7.74\%&44.10&6.63\%&42.60&5.27\%\\
ICT \cite{38}&&&41.64&2.80\%&45.32&9.63\%&46.13&8.66\%&44.36&7.03\%\\	
UAMT \cite{14}&&&39.62&0.78\%&43.77&8.08\%&44.53&7.06\%&42.64&5.31\%\\
CCT \cite{59}&&&42.29&3.45\%&46.50&10.81\%&46.22&8.75\%&45.00&7.67\%\\
DTC \cite{57}&&&36.72&-2.12\%&42.73&7.04\%&42.98&5.51\%&40.81&3.48\%\\
CPS \cite{60}&&&40.03&1.19\%&47.00&11.31\%&48.00&10.53\%&45.01&7.68\%\\
UPRC \cite{56}&&&37.81&-1.03\%&42.30&6.61\%&43.17&5.70\%&41.09&3.76\%\\
PCA(w/o CGAN)&&&\textbf{49.80}&\textbf{10.96\%}&57.61&21.92\%&\textbf{55.40}&\textbf{17.93\%}&\textbf{54.27}&\textbf{16.94\%}\\
PCA(w/ CGAN)&&&44.59&5.75\%&\textbf{57.71}&\textbf{22.02\%}&55.20&17.73\%&52.50&15.17\%\\
\bottomrule 
\end{tabular}
\end{table*}

\subsection{Experiments on Brain Tumor Image Data Set}
We further tested our proposed method on the brain tumor dataset from the public 2019 Brain Tumor Segmentation Challenge. Since the test set and validation set are not publicly available, we randomly divided the training set into the training set and test set at a ratio of 8:2. The input is T1-CE modal brain images, and the output result is the binarized segmentation of the three target regions. On the test set, we performed a quantitative comparison of all semi-supervised segmentation methods on the test set and trained the model with 10\% labeled data.

The results in Table IV show that among these semi-supervised segmentation methods, our PCA (without CGAN) model achieves the best performance (Mean DSC = 52.99\%) in both settings. Compared with the supervised segmentation method trained on only 10\% of labeled data, the improvement is 15.66\%. What's more, we find that unconditional GAN performs much better than conditional GAN. A possible reason could be that small-amplitude jitter produces significant changes in the gray value of brain tumor images. With the introduction of the constraints of the original image information, our model also received noisy information. We visualized the segmentation results in Fig. 5. Compared with other methods, our results have a higher overlap rate with the ground truth, produce fewer false positives, and remain more detailed. This part of the experiment further demonstrates the effectiveness of our method.

\section{Conclusion}
In this paper, we propose a novel semi-supervised learning model (PCA) for medical image segmentation. Specifically, we built a PatchGAN and CGAN framework based on GAN to mine segmentation contextual information and stabilize discriminator training, thereby improving the accuracy of the segmentation model. In addition, we propose a novel conditional input for the GAN framework to alleviate the over-fitting of unlabeled data prediction. Extensive experiments on two public data sets prove the effectiveness of the method. Compared with the current state-of-the-art methods, our proposed model shows superior performance. In the future, we will extend the segmentation scheme to other medical image segmentation tasks that lack enough annotated data and explore the potential of PCA in more visual tasks.
\begin{figure*}[ht]
\includegraphics[width=\textwidth]{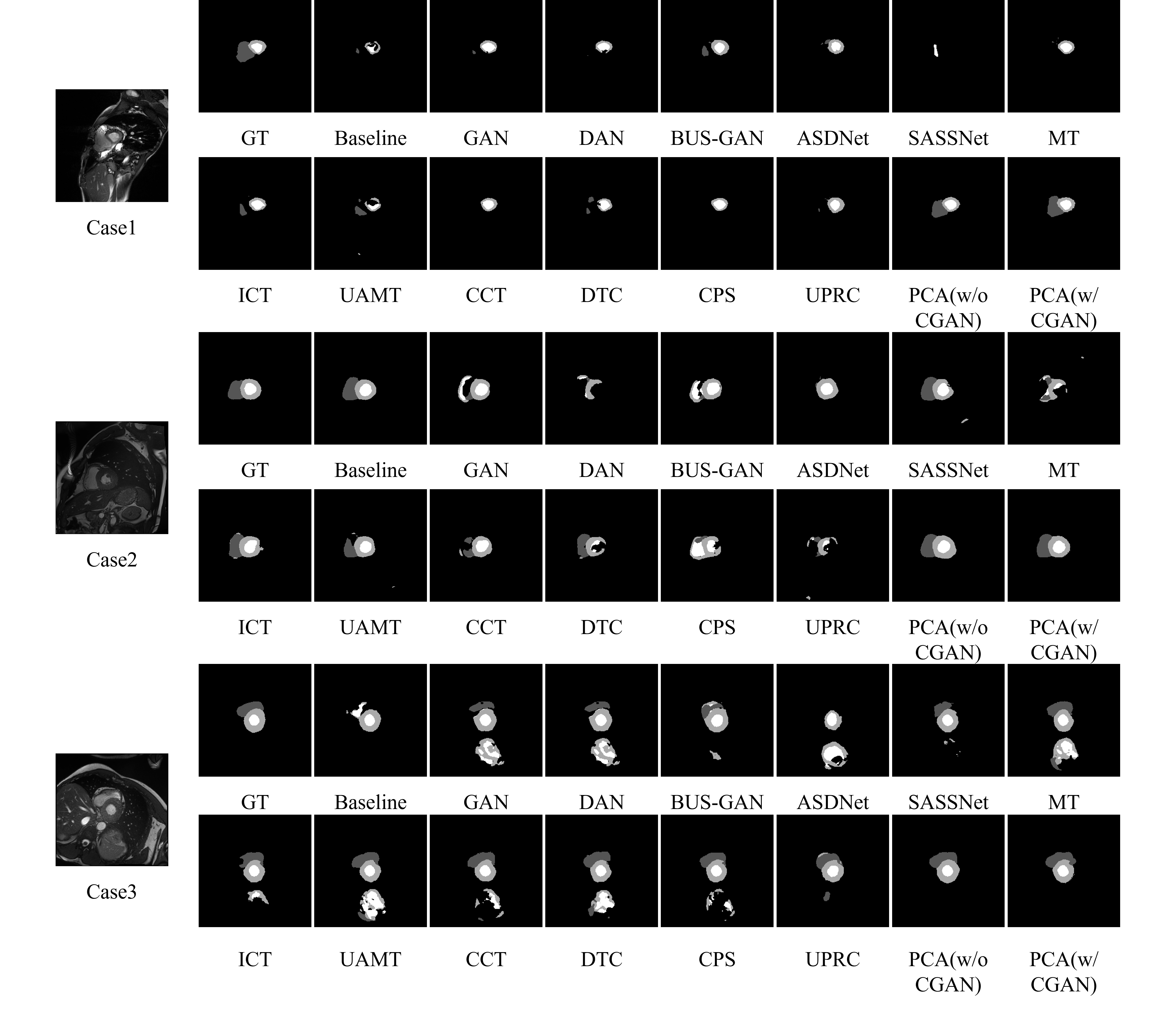}\vspace{-1em}
\caption{The visualization of different semi-supervised segmentation methods under 10\% labeled data.}
\end{figure*}
\begin{figure*}[ht]
\includegraphics[width=\textwidth]{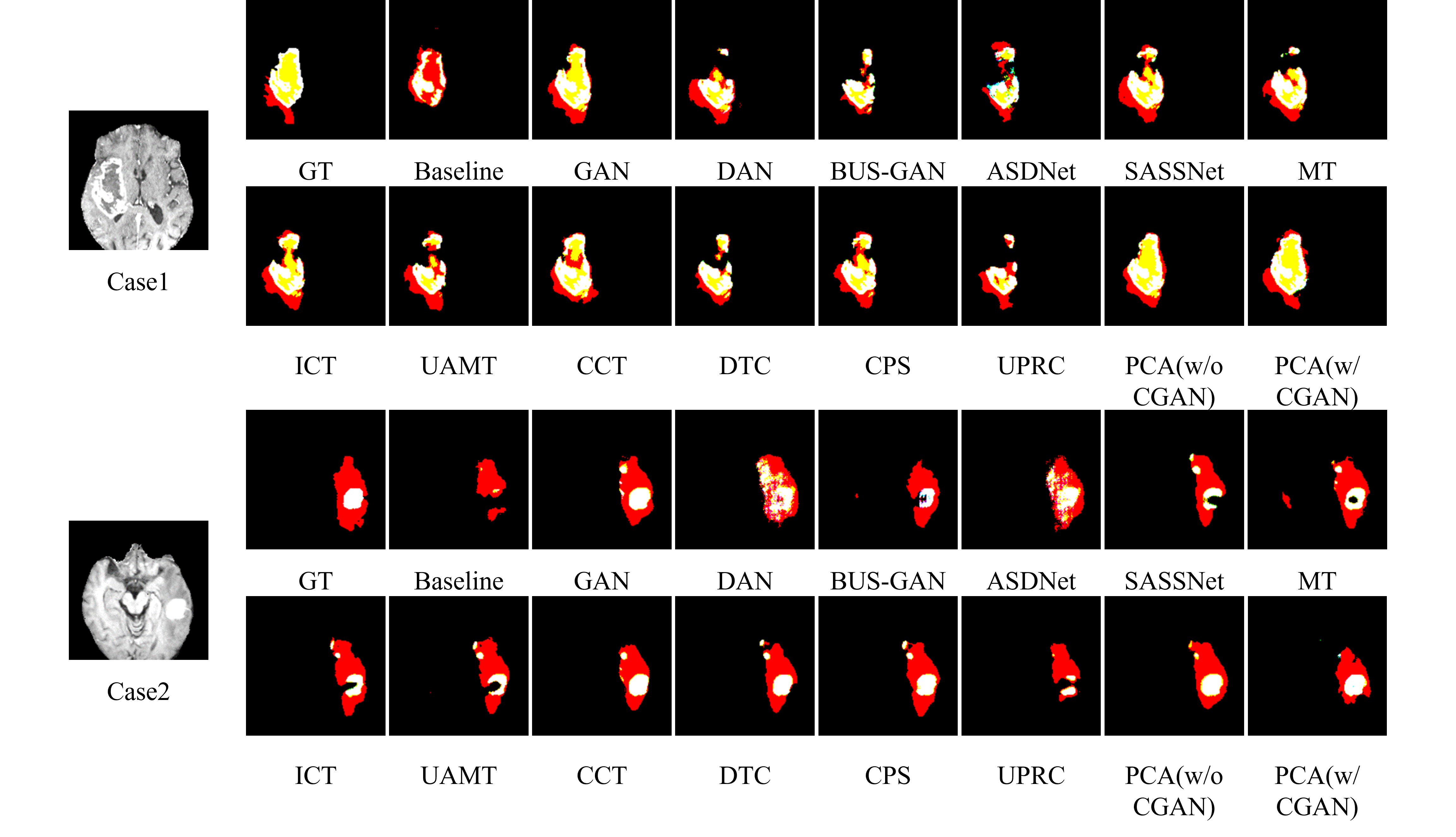}
\caption{The visualization of different semi-supervised segmentation methods under 10\% labeled data. Black corresponds to the background, the green area corresponds to the whole tumor region (wt), the yellow area represents the core tumor region (ct), and the red area represents the enhancing tumor region (et).}\vspace{-1.5em}
\label{fig8}
\end{figure*}
\bibliographystyle{IEEEtran}
\bibliography{draft}

\end{document}